\documentclass[10pt,twocolumn,letterpaper]{article}

\usepackage[pagenumbers]{for_arxiv} 
\definecolor{arxivblue}{rgb}{0.21,0.49,0.74}
\usepackage[pagebackref,breaklinks,colorlinks,allcolors=arxivblue]{hyperref}
\usepackage{cuted}
\usepackage{capt-of}
\usepackage{newfloat}
\usepackage{listings}
\usepackage{bm}  
\usepackage{booktabs, multirow} 
\usepackage{kotex}
\usepackage{amsfonts}
\usepackage{algorithm}
\usepackage{amsmath}
\usepackage{makecell}
\usepackage[table]{xcolor}  
\DeclareCaptionStyle{ruled}{labelfont=normalfont,labelsep=colon,strut=off} 
\floatstyle{ruled}
\newfloat{listing}{tb}{lst}{}
\floatname{listing}{Listing}

\title{Single-step Diffusion for Image Compression at Ultra-Low Bitratess}

\author{%
  Chanung Park \quad
  Joo Chan Lee \quad
  Jong Hwan Ko \\
  Sungkyunkwan University \\
  Suwon, South Korea \\
  {\tt\small pcw980420@g.skku.edu, maincold2@skku.edu, jhko@skku.edu}
}

\begin{document}

\maketitle

\begin{strip}
  \centering
  \includegraphics[width=\linewidth]{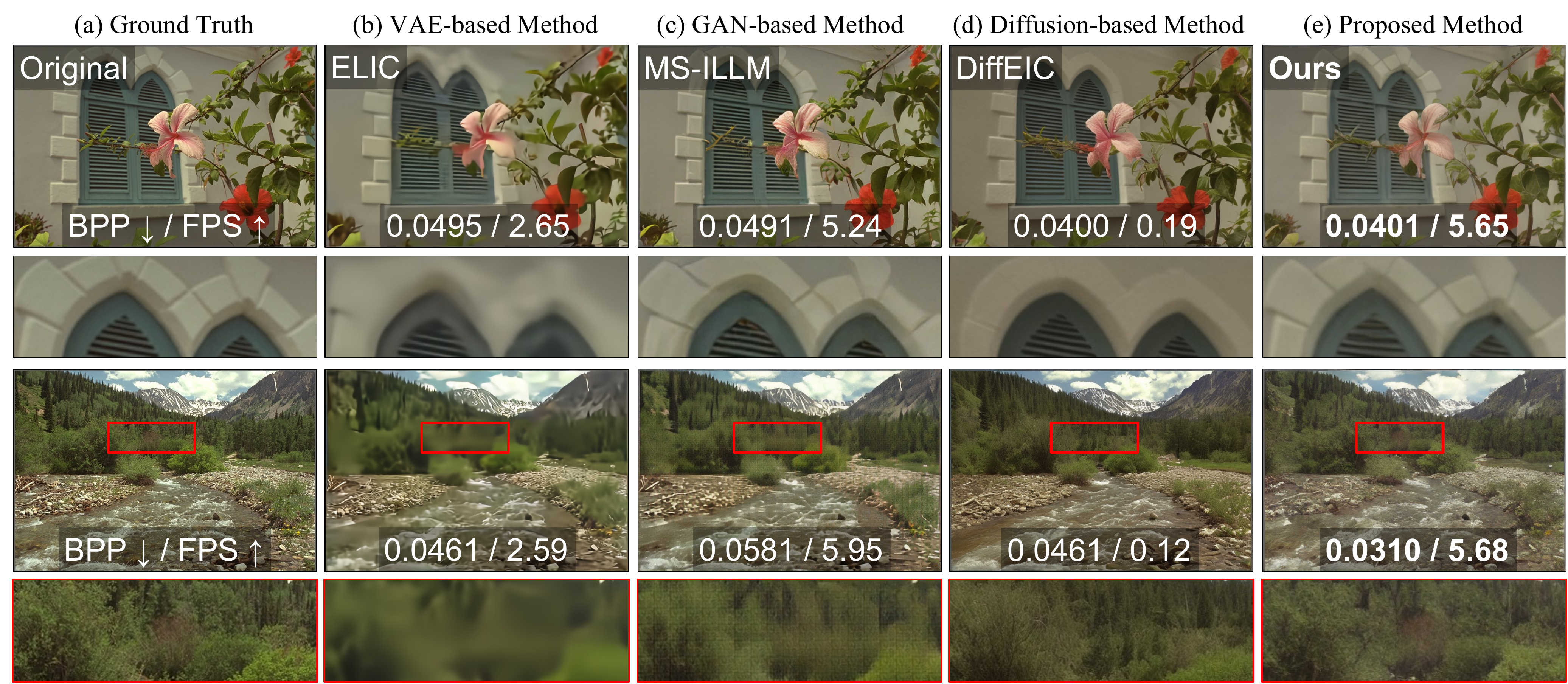}
  \vspace{-1.5em}
  \captionof{figure}{We propose a single-step diffusion method for image compression, which incorporates VQ-Residual training and rate-aware noise modulation. Our approach achieves high perceptual quality and fast decoding at ultra-low bitrates, outperforming state-of-the-art diffusion-based image codecs while enabling about 50× faster decoding.}
  \label{fig:fig1}
\end{strip}

\begin{abstract}
Although there have been significant advancements in image compression techniques, such as standard and learned codecs, these methods still suffer from severe quality degradation at extremely low bits per pixel. While recent diffusion-based models provided enhanced generative performance at low bitrates, they often yields limited perceptual quality and prohibitive decoding latency due to multiple denoising steps.
In this paper, we propose the single-step diffusion model for image compression that delivers high perceptual quality and fast decoding at ultra-low bitrates. Our approach incorporates two key innovations: (i) Vector-Quantized Residual (VQ-Residual) training, which factorizes a structural base code and a learned residual in latent space, capturing both global geometry and high‑frequency details; and (ii) rate‑aware noise modulation, which tunes denoising strength to match the desired bitrate. Extensive experiments show that ours achieves comparable compression performance to state-of-the-art methods while improving decoding speed by about 50× compared to prior diffusion-based methods, greatly enhancing the practicality of generative codecs.
\end{abstract}
    
\vspace{-1.5em}
\section{Introduction}
\label{sec:intro}

Efficient image compression lies at the core of digital communication, storage, and multimedia applications, where minimizing data size while preserving visual quality is essential. Over the past decades, traditional compression algorithms such as JPEG~\cite{jpeg}, JPEG2000~\cite{jpeg2000}, and BPG~\cite{bpg} have been widely adopted, relying on hand-crafted transformations and statistical models to achieve compact representations. With the advent of neural networks, learned image codecs~\cite{ballelic, cheng2020,elic,hyperprior} have been proposed, demonstrating high compression efficiency.

Both the conventional and learnable codecs are typically designed based on information-theoretic principles, where they reduce entropy by selectively discarding visually imperceptible high-frequency components, enabling a more compact and efficient representation of images. However, when the bitrate becomes extremely constrained and the amount of preserved information falls below a certain threshold, these models are incapable of reconstructing the original image, showing deteriorated quality, as illustrated in Fig.~\ref{fig:fig1}(b). This potentially limits the practicality in low-rate scenarios. 

To mitigate this challenge, several approaches have incorporated generative adversarial networks (GANs)~\cite{hific,agustsson2019generative,compressnet,improving,fidelity} to leverage generative capabilities for image reconstruction under ultra-low bitrates, significantly improving perceptual quality of images. Nevertheless, GAN-based codecs are prone to mode collapse and often exhibit unstable texture synthesis, which can be observed in Fig.~\ref{fig:fig1}(c).

More recently, diffusion models~\cite{dpm,ddpm,ddim}, which capable of generating high-quality, high-resolution images through an iterative denoising process, have become a dominant paradigm in generative models. This has motivated recent efforts to leverage diffusion models for image compression tasks, aiming to further enhance perceptual fidelity at low rates (Fig.~\ref{fig:fig1}(d)).
However, despite their impressive generative capabilities, diffusion models often prioritize semantic consistency over fine-grained perceptual details~\cite{ldm}. While they excel at generating semantically coherent and high-resolution images, applying this strength to image compression—where maintaining perceptual similarity to the original input is critical—remains a significant challenge. As a result, diffusion-based image compression methods~\cite{diffc,perco,cdc,diffeic,disney,hoogeboom} typically suffer from limited rate-distortion performance. Furthermore, the inherently iterative nature of the denoising process leads to substantial computational overhead, making these models impractically slow for real-world compression applications.

\begin{figure}[t]
\setlength{\abovecaptionskip}{0pt} 
\setlength{\belowcaptionskip}{0pt} 

\begin{center}
    \includegraphics[width=1.0\linewidth]{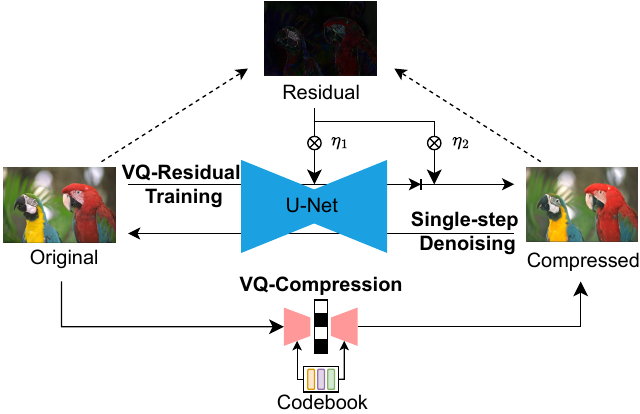}
\end{center}
\vspace{-0.5em}
\caption{
Overview of the proposed method. The input image is first encoded into discrete latent codes via a VQ-compression module with a learnable codebook. During training, the residual between the original image and its VQ-compressed reconstruction is modeled using a U-Net conditioned on the latent code. The U-Net is trained to perform single-step denoising, guided by both the residual signal and semantic prior from the compressed latent.
}

\label{fig:1_1}
\end{figure}

To address these challenges, we propose a novel single-step diffusion model specifically designed for perceptual image compression at ultra-low bitrates. In contrast to conventional approaches that rely on multiple iterative denoising steps, our method reconstructs images in a single step, significantly accelerating the decoding process (Fig.~\ref{fig:fig1}(e)). The key design incorporates vector quantization (VQ) for latent compression, coupled with a single-step residual generation module that learns to recover the difference between compressed and original latents (Fig.~\ref{fig:1_1}), thereby preserving both structural integrity and perceptual quality. Furthermore, we introduce the rate-aware noise modulation mechanism, which adjusts the denoising strength according to the operating bitrate. 


Extensive experiments demonstrate that our method achieves high perceptual rate-distortion performance on par with state-of-the-art image compression methods at ultra-low bitrate. In particular, it outperforms the recent diffusion-based approach DiffEIC~\cite{diffeic} in both visual and perceptual quality, while reducing storage requirements. Furthermore, this compression efficiency is achieved together with over 50× faster decoding enabled by the proposed single-step diffusion incorporated with a lightweight base network (210 M parameters for ours versus 1.4 B for DiffEIC).

\begin{figure*}[t]
\setlength{\abovecaptionskip}{0pt} 
\setlength{\belowcaptionskip}{0pt} 
\begin{center}
    \includegraphics[width=1\linewidth]{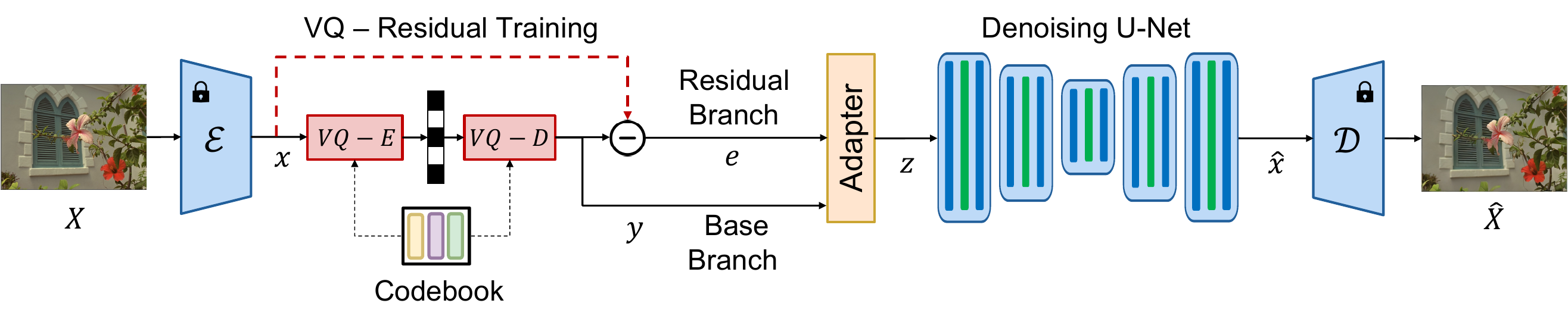} 
\end{center}    
\vspace{-0.5em}
\caption{The proposed framework encodes the input image via a VQ-autoencoder and processes the latent through two branches: a residual branch for structural reconstruction and a base branch for perceptual refinement using a denoising U-Net.}
\label{fig:3}

\end{figure*}

\section{Related Work}
Neural networks have become the foundation of modern lossy image compression codecs, often surpassing traditional methods~\cite{bpg, jpeg, jpeg2000, vvc, hevc} in rate–distortion performance. Early learned approaches employed autoencoder architectures optimized for pixel-wise distortion, typically using variational autoencoders (VAEs) with learned entropy models to compress latent representations~\cite{ballelic,hyperprior,mbt2018,cheng2020}. 
However, distortion-based optimization often leads to overly smooth reconstructions, especially at ultra-low bitrates.
\subsection{Image Compression at Ultra-low Bitrates}
To enhance perceptual quality at ultra-low bitrates, adversarial and perceptual losses have been integrated. GAN-based codec were introduced to enable perceptual image compression at low-bitrates~\cite{agustsson2019generative}, demonstrating that realistic textures can be reconstructed from highly compressed codes.  HiFiC~\cite{hific} further advanced this direction, using a GAN and perceptual loss to achieve high-fidelity reconstructions. ILLM~\cite{msillm} introduced implicit local likelihood models that improve texture detail via localized modeling of statistical fidelity. GLC~\cite{glc} presented latent coding for generative reconstruction. 

\vspace{-1.0em}
\paragraph{Diffusion-based Image Compression.}
Diffusion models have recently emerged as highly capable generative models, surpassing GANs in perceptual quality. CDC~\cite{cdc} proposed a conditional diffusion-based codec, using compact latents from a VAE encoder to guide image reconstruction. This method demonstrated superior perceptual quality compared to GAN-based decoders at low-bitrates. More recent methods utilize pre-trained latent diffusion models (LDMs)~\cite{ldm}. PerCo~\cite{perco} has proposed LDM-based perceptual compression, conditioning on both vector-quantized latents and textual descriptions. DiffEIC~\cite{diffeic} and DiffPC~\cite{diffpc} have combined compressive VAEs with pre-trained diffusion models to reconstruct realistic images at ultra-low bitrates. HDCompression~\cite{hdc} proposed a hybrid approach that intergrates diffusion models with conventional codecs. RDEIC~\cite{rdeic} further improves performance with less denoising steps by introducing a relay residual strategy. Although diffusion models have demonstrated superior performance compared to GAN-based methods, they still suffer from significant computational overhead, as they require either executing dozens of denoising steps or employing large-scale neural networks with substantial parameter complexity.

\subsection{Acceleration of Diffusion Models}
Due to a major limitation of diffusion models—their inherently iterative nature~\cite{ddpm}—several strategies have been proposed to accelerate inference. DDIM~\cite{ddim} and DPM-Solver~\cite{dpm} reduce sampling time by skipping intermediate steps or solving an ordinary differential equation (ODE) approximation of the reverse process, often generating high-fidelity samples in 10–20 steps. Salimans and Ho \textit{et al.} ~\cite{salimans} proposed progressive distillation, iteratively halving the number of diffusion steps. DMD ~\cite{dmd} introduced distribution matching distillation , which trains a single-step generator to match the output distribution of a full diffusion model, enabling real-time generation at high perceptual quality. Consistency models enforce consistency constraints across noise levels and allow single-step inference~\cite{consistency}. EDM ~\cite{edm} introduces elucidating the design space for specific design choices.


Separate from these acceleration techniques that have primarily focused on unconditional or class-conditional generation, recent work has also explored more intricate tasks of image reconstruction.
For instance, ResShift~\cite{resshift} introduced an efficient diffusion method for image super-resolution. Rather than modeling a standard forward noise process, they proposed a residual-guided Markov chain based on the difference between the high-resolution (HR) image and its low-resolution (LR) input. Specifically, HR $x_0$ is gradually shifted toward the LR $y_0$ during T time steps with residual $e_0 = x_0 - y_0$. The transition function is formulated as,
\vspace{-0.5em}
\begin{equation}
\begin{aligned}
q(x_t \mid x_0, y_0) &= \mathcal{N}\Big(x_t\ ;\ x_0 + \eta_t e_0, 
\kappa^2 \eta_t\, \mathbf{I} \Big),
\end{aligned}
\end{equation}
where $\eta_t$ is a time-dependent shift factor and $\kappa$ controls the overall noise variance with $t$ uniformly sampled from $\{1,\cdots,T\}$. Conversely, the reverse process from $y_0$ to $x_0$ can be formulated as follows,
\begin{equation}
\begin{aligned}
p_\theta(x_{t-1} \mid x_t, y_0) 
\\&\!\!\!\!\!\!\!\!\!\!\!\!\!\!\!\!\!\!\!\!\!\!\!=\mathcal{N} \Big( x_{t-1}\ \big|\ \mu_\theta(x_t, y_0, t),\ 
\kappa^2 \frac{\eta_{t-1}}{\eta_t} \alpha_t\, \mathbf{I} \Big)
\end{aligned}
\label{eq:resshift}
\end{equation}
\begin{equation}
\mu_\theta(x_t, y_0, t) = \frac{\eta_{t-1}}{\eta_t} x_t + \frac{\alpha_t}{\eta_t} f_\theta(x_t, y_0, t),
\label{eq:resshift_rev}
\end{equation}
where $\alpha_t=\eta_t-\eta_{t-1}$, and the estimated mean is parameterized by a neural network $f_\theta$. This approach reduces the number of diffusion steps to 15, enabling much faster inference while maintaining quality.

\begin{figure*}[t]
\setlength{\abovecaptionskip}{0pt} 
\setlength{\belowcaptionskip}{0pt} 
\begin{center}
    \includegraphics[width=1.0\linewidth]{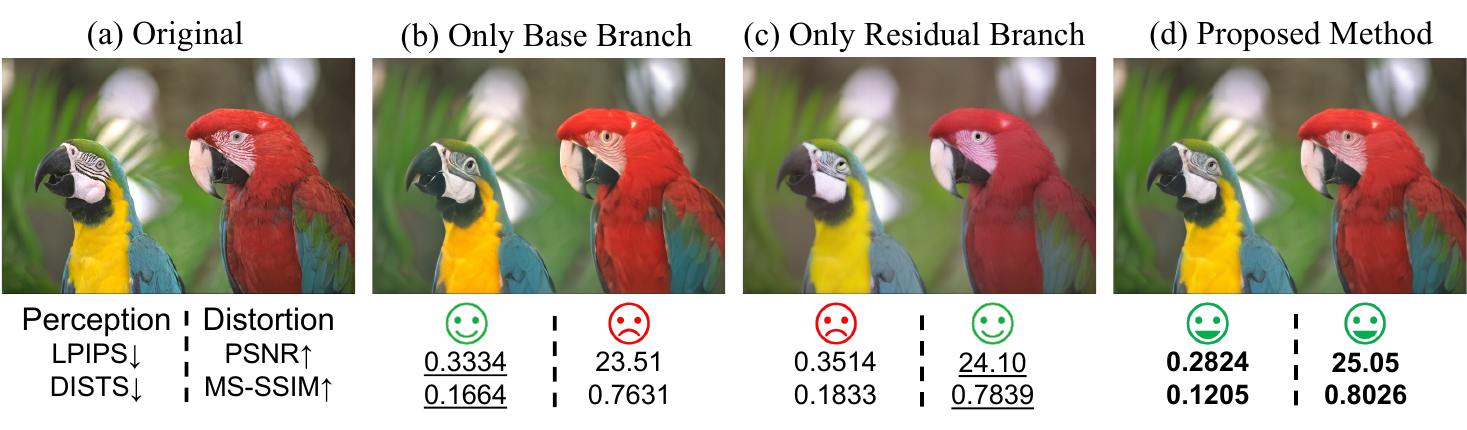} 
\end{center}
\vspace{-1.0em}
\caption{Reconstruction comparison. (a) Original image. (b) Proposed method combining base+residual branches achieves the best perceptual/distortion quality. (c) Base-only branch using VQ-based structure reconstruction without refinement. (d) Residual-only branch yields higher distortion scores (PSNR, MS-SSIM) but lacks perceptual quality (LPIPS, DISTS).}
\vspace{-1.0em}
\label{fig:4}
\end{figure*}

\section{Method}
The overall pipeline of our proposed perceptual image compression framework is depicted in Fig.~\ref{fig:3}. 
The input RGB image $X\in \mathbb{R}^{H\times W\times 3}$ is initially encoded into a latent representation $x =\mathcal{E}(X)$ with an encoder $\mathcal{E}$.
After compression and a single denoising process, the resulting feature $\hat{x}$ is decoded back to the output RGB image $\hat{X}=\mathcal{D}(\hat{x})$ using a decoder $\mathcal{D}$. 

\subsection{Latent Feature Compression}
In contrast to approaches such as DiffEIC~\cite{diffeic} or RDEIC~\cite{rdeic}, which rely on entropy coding or partially incorporate vector quantization (VQ), our method constructs the bitstream entirely through VQ. The encoder produces latent features $x \in \mathbb{R}^{h\times w\times d}$, where $h$, $w$, and $d$ denote the height, width, and channel dimension, respectively. These features are then discretized using a learned codebook $\mathcal{V}={v[k]}^K_{k=1} \subset \mathbb{R}^d$~\cite{vqvae}, enabling compact and efficient representation.
Each vector $x_{i,j}\in\mathbb{R}^d$ at spatial position $i,j$ is quantized to their nearest codebook entry, and the resulting quantized feature $q \in \mathbb{Z}^{h\times w\times d}$ and compressed feature $y \in \mathbb{R}^{h\times w\times d}$ are formulated as follows:

\begin{align}
q_{i,j} = \arg\min_k || x_{i,j} - v[k]||^2_2, \,\,\,\,
y_{i,j} = v[q_{i,j}]
\end{align}

Note that, in contrast to representation learning frameworks such as VQGAN~\cite{vqgan}, we adjust the codebook size and latent resolution based on the target bits-per-pixel (BPP) to enable extreme compression through vector quantization and arithmetic coding. The actual BPP $B$ is measured as:
\begin{equation}
B = \frac{1}{H  W} \sum_{i,j} -\log_2 \text{PMF}(y_{i,j}),
\end{equation}
where $\text{PMF}(\cdot)$ denotes the probability mass function.

This makes the rate directly dependent on the quantized indices, thereby eliminating the large variability of bitrate commonly observed in entropy-coded representations. Consequently, the actual output bitrate closely aligns with the target bitrate with minimal variance across different images, as further demonstrated in experiments ~\ref{sec:4.2}.

\subsection{Single-step Denoising for Image Compression}
To accelerate diffusion-based image compression, adopting techniques such as ResShift, which has demonstrated promising results in super-resolution, may serve as a viable and effective option. However, directly applying it to image compression imposes unique challenges. 
In super-resolution, the inputs to the diffusion model retain sufficient perceptual information, which allows diffusion models to learn how to generate high-frequency details. 
In contrast, image compression inherently involves heavily compressed inputs, where perceptual details are often the first to be lost (even before semantic components) due to the nature of the compression process. This makes it particularly difficult for diffusion models to accurately reconstruct fine-grained visual information~\cite{ldm}.
As a result, performing a large number of denoising steps often degrades perceptual quality rather than improving it (Table ~\ref{tab:step}). This has led recent approaches to adopt lightweight diffusion models with only a few steps (e.g., two) for low-bit image compression~\cite{rdeic}.

Leveraging this insight, we propose using only a single forward denoising step, which is enough to exploit generative ability of diffusion models for high perceptual quality, while enabling fast inference by avoiding the repeated steps.
Thus, Eq.~\ref{eq:resshift} can be simplified as follows:

\begin{equation}
    q(\tilde{x} \mid x, y) = \mathcal{N}( \tilde{x};\, x + \eta_q(y-x),\ \kappa^2 \eta_q \mathbf{I}),
\end{equation}
\begin{equation}
    p_\theta(\hat{x} \mid \tilde{x}, y) = \mathcal{N}( \hat{x}\ |\ f_\theta(\tilde{x}, y),\ \kappa^2 \eta_p \mathbf{I}) 
\end{equation}

where $\tilde{x}$ is the noise-imposed feature obtained by the forward pass from $x$, and $\hat{x}$ is the denoised feature produced by the reverse pass from $y$. $\eta_q, \eta_p$ are the noise scale for the forward and reverse processes.

Although directly training a single-step model can achieve satisfactory results, we improve the training stability and generalizability by incorporating an additional step with minimal noise only for the training phase. In other words, we train our model with 2-step diffusion (following Eq.~\ref{eq:resshift}): one step for perceptual denoising with a large-scale noise, which is the only step used for inference, and the other for distortion robust training with a tiny noise. This strategy ensures robust training with high-quality image reconstruction while maintaining single-step fast inference (see Table~\ref{tab:step}).

\paragraph{Residual Fusion U-Net}
Although denoising from the compressed latent retains semantic content relatively well, it tends to introduce artifacts at the pixel level (Fig.~\ref{fig:4} (a)). To address this limitation and ensure high-fidelity reconstruction, we propose a residual fusion strategy.
Specifically, the compressed latent $y$ undergoes parallel processing via two distinct, yet complementary branches of the base $y$ and the residual $e=\tilde{x}-y$. The base branch focuses on the low-level contexts. In contrast, the residual branch is responsible for capturing and reconstructing the high-level structural content of the image. The outputs from the residual and base branches are fused via a latent adapter $A(\cdot)$, producing an integrated representation $z$. Subsequently, the U-Net $U(\cdot)$ refines the residual between the original latent representation and the semantic reconstruction, focusing on recovering fine-grained details. Through this denoising process, the base branch effectively suppresses perceptual artifacts and produces a more perceptually faithful latent representation.
Formally, the decoded output can be written as follows,
\begin{equation}
    f_\theta(\tilde{x},y) = U(z),
\end{equation}
\begin{equation}
    z := 
A(\tilde{x},y) = \text{conv}(\text{concat}(\tilde{x}-y, y)),
\end{equation}

where $\text{conv}(\cdot)$ and $\text{concat}(\cdot,\cdot)$ are a single convolution layer and concatenation function, respectively.

As shown in Fig.~\ref{fig:4}, this fusion results in a harmonized latent feature  that captures both structural fidelity and perceptual refinement. Finally, the adapted latent  is passed through the decoder  to reconstruct the final output image by $\hat{X}=\mathcal{D}(\hat{x})$. This architectural design strategically balances compression efficiency, semantic accuracy, and perceptual quality, ensuring optimal performance especially under ultra-low bitrate conditions.

The model is trained in an end-to-end manner by minimizing the loss, defined as a weighted sum of multiple objectives: a reconstruction loss in pixel space, a perceptual loss measured by LPIPS to enhance visual quality, and two structural losses applied to the semantic and compressed representations, can be formulated as:
\begin{equation}
\begin{aligned}
L=\|X-\hat{X}\|_2^2 + \lambda\mathcal{L}_{lpips}(X,\hat{X})  
\\&\!\!\!\!\!\!\!\!\!\!\!\!\!\!\!\!\!\!\!\!\!\!\!\!\!\!\!\!\!\!\!\!\!\!\!\!\!\!\!\!\!\!\!\!\!\!\!\!\!\!+ \|\text{sg}(x)-y\|_2^2 +\beta\|\text{sg}(y)-x\|_2^2,
\label{eq:loss}
\end{aligned}
\end{equation}
where $\text{sg}(\cdot)$ is the stop-gradient operator.


\begin{figure}
\setlength{\abovecaptionskip}{0pt} 
\setlength{\belowcaptionskip}{0pt} 
\begin{center}
    \includegraphics[width=1.00\linewidth]{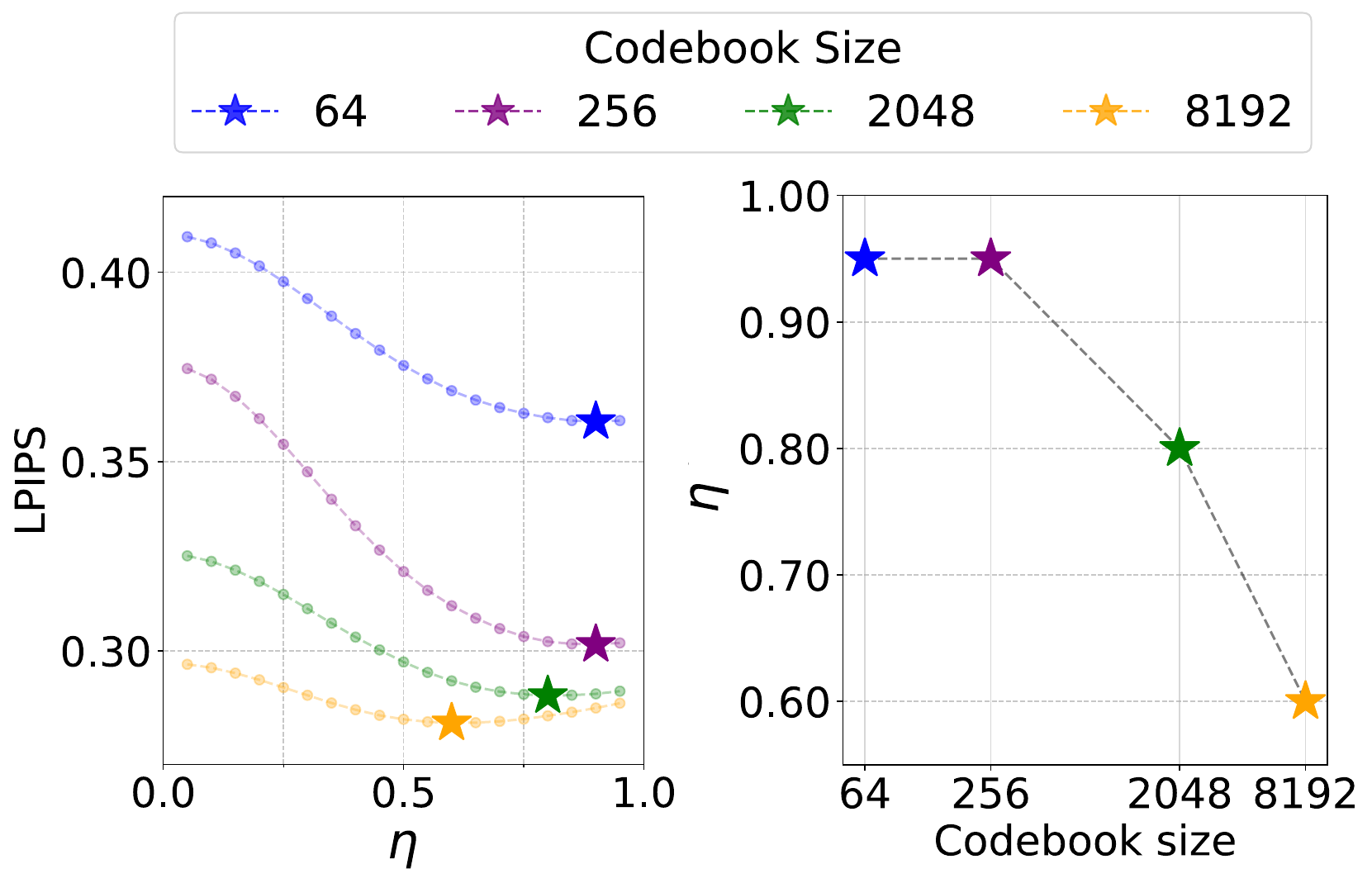}
\end{center}
\vspace{-1.0em}
\caption{
(Left) LPIPS vs. noise modulation parameter $\eta$ for different codebook sizes (64–8192); larger codebooks correspond to higher bitrates. (Right) As codebook size increases, the optimal $\eta^*$ minimizing LPIPS shifts lower, suggesting weaker denoising is needed at higher bitrates due to reduced quantization error.
}
\label{fig:5}
\end{figure}

\subsection{Rate-aware Noise Modulation}
As shown by Li \textit{et al.}~\cite{diffeic} and Relic \textit{et al.}~\cite{disney}, lower bitrates typically require more denoising steps to achieve high perceptual quality in reconstruction. This is because the noise schedule in conventional diffusion models is fixed regardless of the bitrate, which necessitates adjusting the number of denoising steps dynamically to accommodate different levels of compression. However, this leads to increased computational cost and significantly slower decoding speed during inference. To overcome this limitation, we propose a rate-aware noise modulation at the single inference step.

\begin{figure*}
\begin{center}
    \includegraphics[width=1\linewidth]{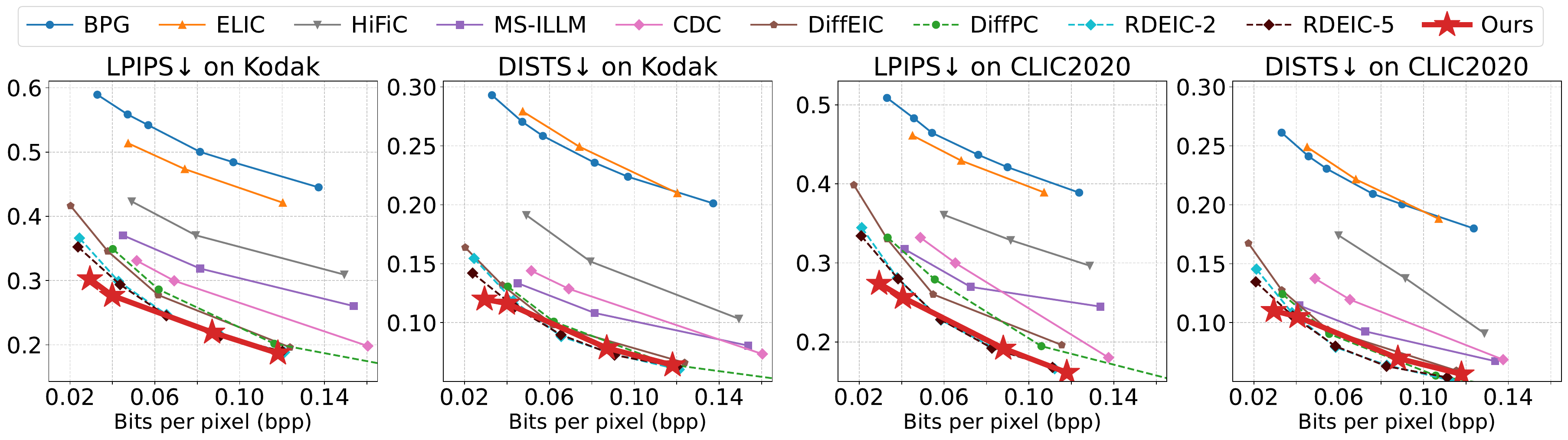}
\end{center}
\vspace{-1.0em}
\caption{Comparison of LPIPS and DISTS across various
methods on the Kodak datasets}
\label{fig:6}

\end{figure*}
From Fig.~\ref{fig:5}, we can observe that as the size of the VQ-codebook increases (i.e., as the BPP increases), the optimal $\eta_q$ value for achieving the best LPIPS score decreases. This empirical trend suggests a negative correlation between optimal $\eta_q$ and BPP, which can be represented as:
\begin{equation}
\eta_q \propto \frac{1}{B}
\end{equation}
This implies that the noise modulation strength $\eta$ should be adjusted according to the bitrate to achieve optimal denoising strength and perceptual quality. Based on this relationship, we adopt a bitrate-dependent noise modulation strategy that adjusts $\eta_q$ at inference time to ensure optimal trade-offs between perceptual quality and decoding efficiency. For instance, at lower bitrates, where the input contains less information due to aggressive quantization, we inject stronger noise (larger $\eta$) during the reverse diffusion step. This allows the model to perform a stronger one-step correction, effectively compensating for the loss of detail without requiring multiple denoising iterations. 

\section{Experiments}

\subsection{Experimental Setup}
For training, we use the ImageNet dataset~\cite{imagenet} with random cropping to a resolution of 256$\times$256. For evaluation, we test on two standard benchmarks: Kodak~\cite{kodak} and CLIC2020~\cite{clic2020}. Following the existing benchmark for testing CLIC2020, each image is resized so that its shorter side is 768 pixels, followed by a center crop to maintain consistency across samples.  We compute LPIPS~\cite{lpips} using a VGG-based backbone with normalized activations, identical to the configuration used during training. DISTS~\cite{dists} is computed using the pretrained metric from the PyIQA library. 
More implementation details are detailed in the supplementary materials.

We compare our proposed method with several representative codecs under ultra-low bitrate settings, including BPG~\cite{bpg}, ELIC~\cite{elic}, HiFiC~\cite{hific}, MS-ILLM~\cite{msillm}, CDC~\cite{cdc}, DiffEIC~\cite{diffeic}, DiffPC~\cite{diffpc}, and RDEIC~\cite{rdeic}. For fair comparison, we reproduced all methods using official implementations when available, except for DiffPC, whose results are reported based on numbers from the original paper due to the lack of publicly available code.

\subsection{Comparisons with State-of-the-art methods}
\paragraph{Quantitative Results}
Fig.~\ref{fig:6} shows the rate-distortion (RD) curves using LPIPS and DISTS as perceptual quality metrics on the Kodak and CLIC2020 datasets. 
Our method consistently outperforms all baselines at ultra-low bitrates ($<$0.05 bpp), delivering substantially better perceptual scores. Across other bitrate ranges, we also achieve comparable perceptual quality with other baseline methods.
Comparisons on traditional distortion metrics such as PSNR and MS-SSIM~\cite{ms-ssim} are provided in the Fig.~\ref{fig:psnr_msssim}.

\begin{table}
\centering
\scriptsize
\renewcommand{\arraystretch}{0.95}
\setlength{\tabcolsep}{3pt}
\resizebox{1.0\linewidth}{!}{
\begin{tabular}{
    c  
    c  
    c  
    cc 
    cc 
}
\toprule
\addlinespace[-1.5ex]
\raisebox{-5.5ex}{\textbf{Model}} & \raisebox{-5.5ex}{\textbf{\#Params}} & \multicolumn{2}{c}{\raisebox{-3.5ex}{\textbf{BD-Rate (\%) ↓}}} & \multicolumn{2}{c}{\raisebox{-3.5ex}{\textbf{Time (Sec) ↓}}} \\
\cmidrule(lr){3-6}
& & LPIPS & DISTS & Encoding & Decoding \\
\midrule      
             ELIC       & 36M      & -         & -          & 0.395 & 0.447  \\
             HiFiC      & 182M     & 132.14    & 52.68      & 0.262 & 0.412  \\
             MS-ILLM    & 182M     & 46.89     & -14.77     & 0.245 & \textbf{0.234} \\
\midrule
             CDC        & 68M      & 0         & 0          & \textbf{0.038} & 10.7428 \\
            DiffEIC    & 1.4B     & -25.22    & -43.04     & 0.801 & 12.502  \\
            DiffPC     & -        & -21.83    & -41.72     & $>$0.089     & $>$7.325 \\
            RDEIC-2    & 1.4B     & -37.86    & -47.83     & 0.939 & 0.548  \\
            RDEIC-5    & 1.4B     & \underline{-39.54} & \textbf{-50.84} & 0.965 & 1.248  \\
            
            \cellcolor{gray!15}Ours       & \cellcolor{gray!15}210M     & \cellcolor{gray!15}\textbf{-45.65} & \cellcolor{gray!15}\underline{-48.23} & \cellcolor{gray!15}\underline{0.136} & \cellcolor{gray!15}\underline{0.253} \\
\bottomrule
\end{tabular}
}
\caption{Comparison of methods in terms of BD-Rate (on Kodak) and encoding/decoding time (on CLIC2020), using the NVIDIA TITAN RTX. The upper three methods and the lower ones are VAE- and diffusion-based approaches, respectively.}
\vspace{-1em}
\label{tab:bd_time}

\end{table}

\begin{figure*}
\begin{center}
    \includegraphics[width=1\linewidth]{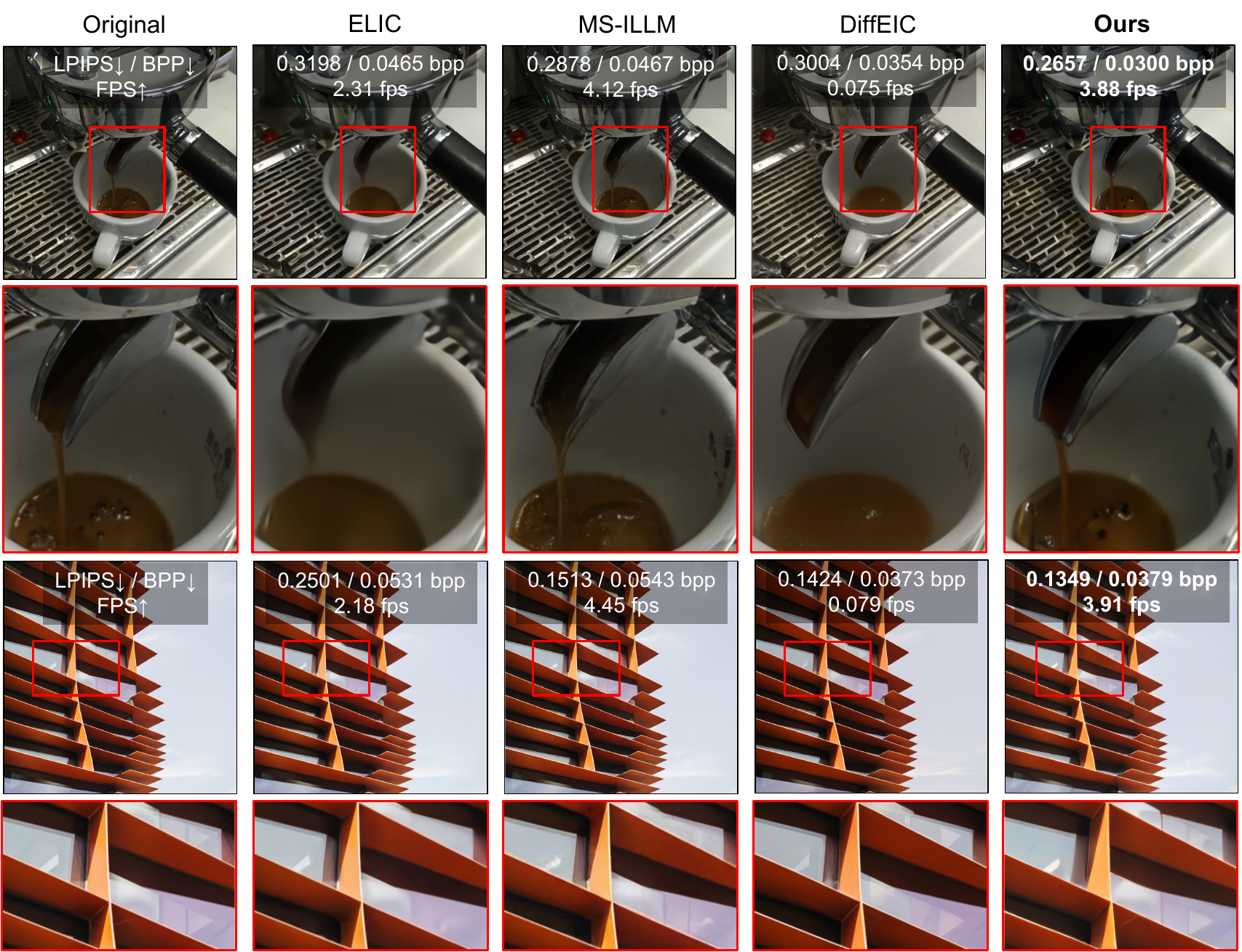} 
\end{center}
\vspace{-1.0em}
\caption{Qualitative examples on the CLIC2020 dataset.}
\label{fig:7}
\vspace{-1.em}
\end{figure*}

\begin{figure}[t]

\begin{center}
    \includegraphics[width=1.00\linewidth]{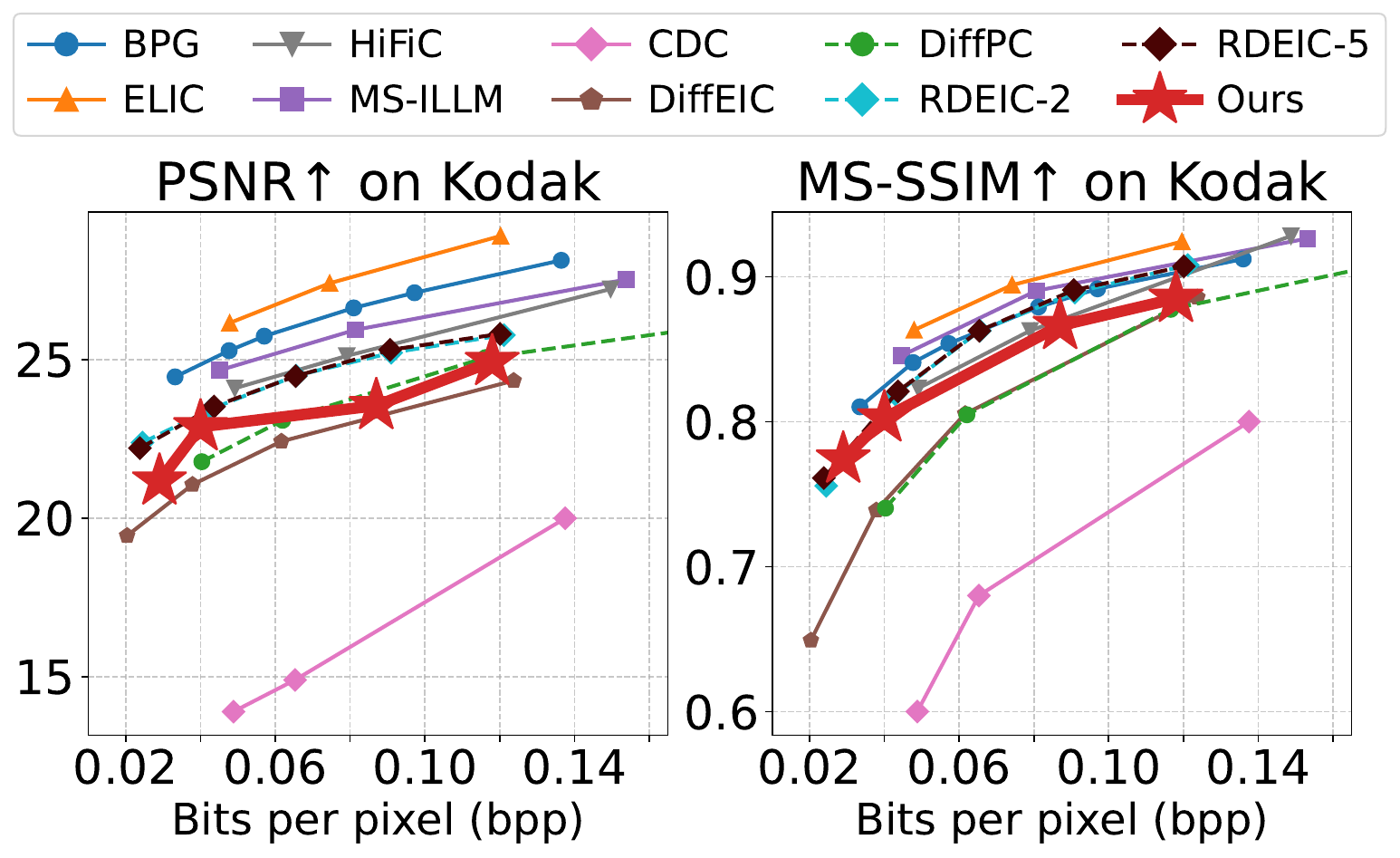} 
\end{center}
\vspace{-1.0em}
\caption{
Comparison of PSNR and MS-SSIM across various methods on the Kodak datasets
}
\label{fig:psnr_msssim}
\end{figure}

\paragraph{Qualitative Results}
Fig.~\ref{fig:fig1} and Fig.~\ref{fig:7} provide visual comparisons of reconstructed images.
VAE-based methods such as ELIC tend to produce overly smoothed results, often losing fine textures and edge details.
In contrast, when compared to MS-ILLM (GAN-based), DiffEIC, and our method (both diffusion-based) yields images more faithful to the original, preserving richer structural and perceptual content, even at lower or similar bitrates.
\paragraph{Complexity Analysis}
Table~\ref{tab:bd_time} reports the BD-Rate~\cite{bd-rate} and the encoding/decoding times of each method.
Our method achieves the lowest BD-Rate in LPIPS and the second lowest in DISTS, indicating better rate-distortion efficiency.
Moreover, compared to the previous diffusion-based method DiffEIC, and our method achieves over 50$\times$ faster decoding speed, demonstrating the effectiveness of our single-step denoising framework in practical deployment scenarios.

\begin{table}
\centering
\resizebox{0.475\textwidth}{!}{
\begin{tabular}{lcc|cc}
\toprule
\multirow{2}{*}{\textbf{Method}} & \multicolumn{2}{c}{\textbf{Perceptual quality ↓}} & \multicolumn{2}{c}{\textbf{Distortion ↑}} \\
\cmidrule(lr){2-5}
& LPIPS & DISTS & PSNR (dB) & MS-SSIM \\
\midrule
1 step      & \underline{0.312}   & \underline{0.127}   & 22.03 & 0.752\\
2 step    & 0.316   & 0.130   & \textbf{22.21} & \underline{0.758}\\
5 step     & 0.317   & 0.129   & 22.17 & 0.753\\
15 step   & 0.326   & 0.138   & 22.15 & 0.751\\
30 step   & 0.323   & 0.136   & 22.10 & 0.749\\
50 step  & 0.319   & 0.134  & 21.94 & 0.742\\

\cellcolor{gray!15}Ours      & \cellcolor{gray!15}\textbf{0.309}   & \cellcolor{gray!15}\textbf{0.122}   & \cellcolor{gray!15}\underline{22.19} & \cellcolor{gray!15}\textbf{0.767}\\
\bottomrule
\end{tabular}}

\caption{Comparison of perceptual and distortion metrics at 0.0294 bpp across different denoising steps. Our method achieves the best perceptual quality (LPIPS, DISTS)  while preserving distortion scores (PSNR, MS-SSIM). Similar trends hold across other bitrates.}
\label{tab:step}
\vspace{-1.0em}
\end{table}

\subsection{Bitrate-Step Analysis}
\label{sec:4.2}
We analyze the relationship between the variance of the actual output bitrate and the corresponding optimal number of diffusion steps.

\begin{figure}
\begin{center}
    \includegraphics[width=1.0\linewidth]{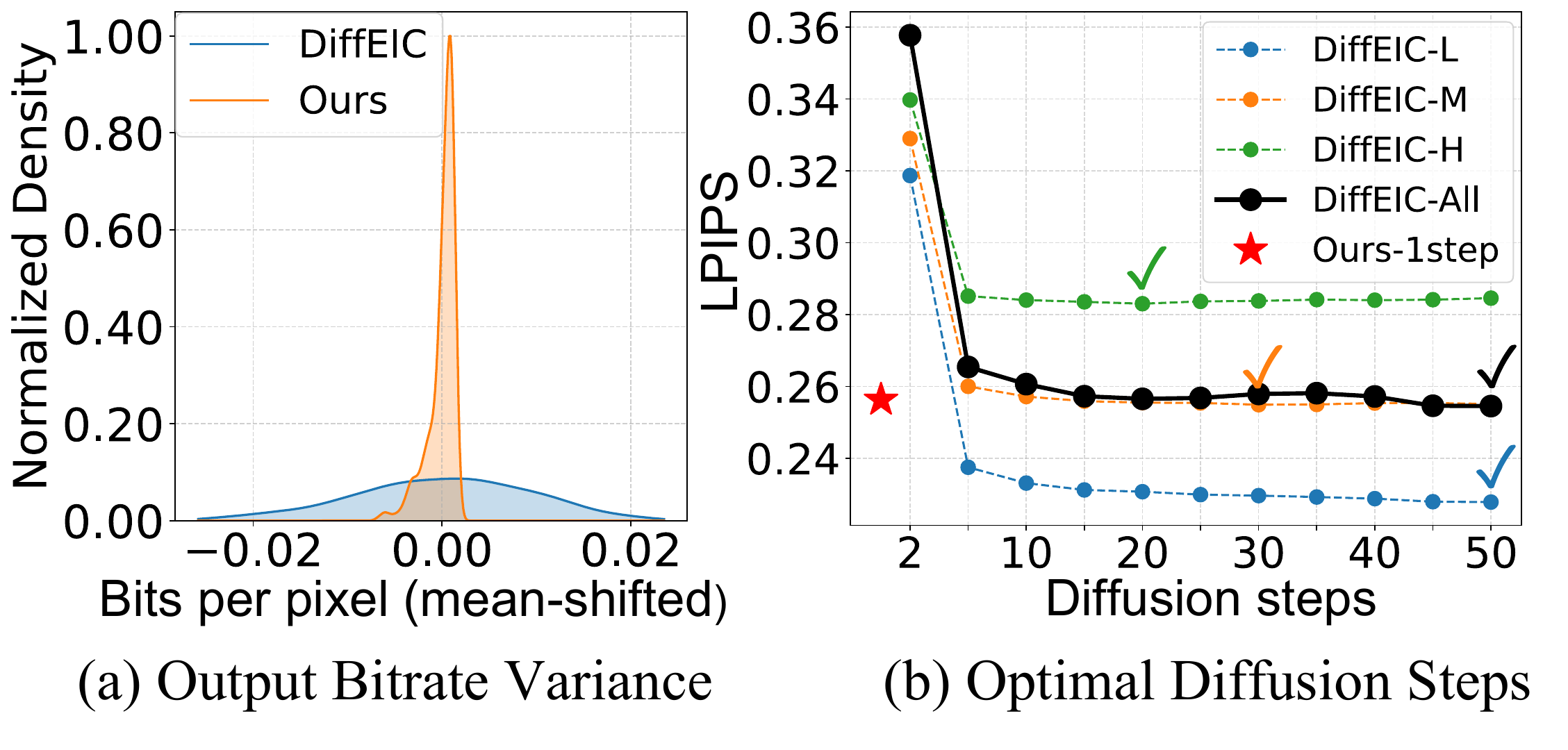} 
\end{center}
\vspace{-1.0em}
\caption{(a) Analysis of output bitrate variance. Compared to DiffEIC, our method achieves tightly bounded output bitrate due to VQ-based compression, resulting in significantly reduced variance across images. (b) LPIPS performance versus diffusion steps. DiffEIC-L, M, H denote subsets of images grouped by their output bitrates—low, medium, and high, respectively—though all are produced by the same model, while DiffEIC-All represents the averaged performance across these subsets. The check marks indicate the optimal diffusion steps for each subset, i.e., the points where the best LPIPS performance is achieved.}
\label{fig:7_2}
\end{figure}

\noindent\textbf{Predictable output bitrate} As shown in Fig.~\ref{fig:7_2} (a), DiffEIC shows significantly higher variance in output bitrate than our method, despite having the same target bitrate. This variance difference arises because DiffEIC relies on entropy coding, where the actual output bitrate fluctuates significantly with image complexity. 
Such high variance complicates codec deployment in bandwidth-limited practical scenarios (e.g., in wireless systems).
In contrast, our method substantially reduces this variance by employing VQ-based compression, which tightly bounds the output bitrate in Fig.~\ref{fig:7_2} (a).
This property ensures that the output bpp is accurately predictable, which is particularly advantageous in practical scenarios where communication protocols must operate under bandwidth constraints.

\noindent\textbf{Adaptive single-step diffusion} Furthermore, diffusion-based approaches such as DiffEIC cause different optimal diffusion steps across input images, making it necessary to adjust the step count accordingly.
As shown in Fig.~\ref{fig:7_2} (b), results of DiffEIC-L, DiffEIC-M, and DiffEIC-H are derived from the same model trained at a single target bitrate, where images in the dataset fall into lower, middle, or higher output bitrate ranges depending on their content complexity. We observe that images in the low-bitrate range (DiffEIC-L) achieve the best LPIPS performance at around 20 steps, those in the middle-bitrate range (DiffEIC-M) at 30 steps, those in the high-bitrate range (DiffEIC-H) at 50 steps. 
This trend indicates that the optimal number of steps increases as the bitrate decreases. 

Although the optimal number of steps varies between 20 and 50, DiffEIC-All achieves its best performance at 50 steps. This incurs unnecessary additional steps for less complex images and may lead to suboptimal compression performance.
In contrast, our method employs a consistent single-step denoising process, with the noise level adaptively adjusted according to the target bitrate. This not only simplifies the inference pipeline but also yields superior performance over DiffEIC at the same target bitrate (Fig.~\ref{fig:7_2} (b)).


\subsection{Ablation Study}
\begin{figure}[t]
\begin{center}
    \includegraphics[width=1.0\linewidth]{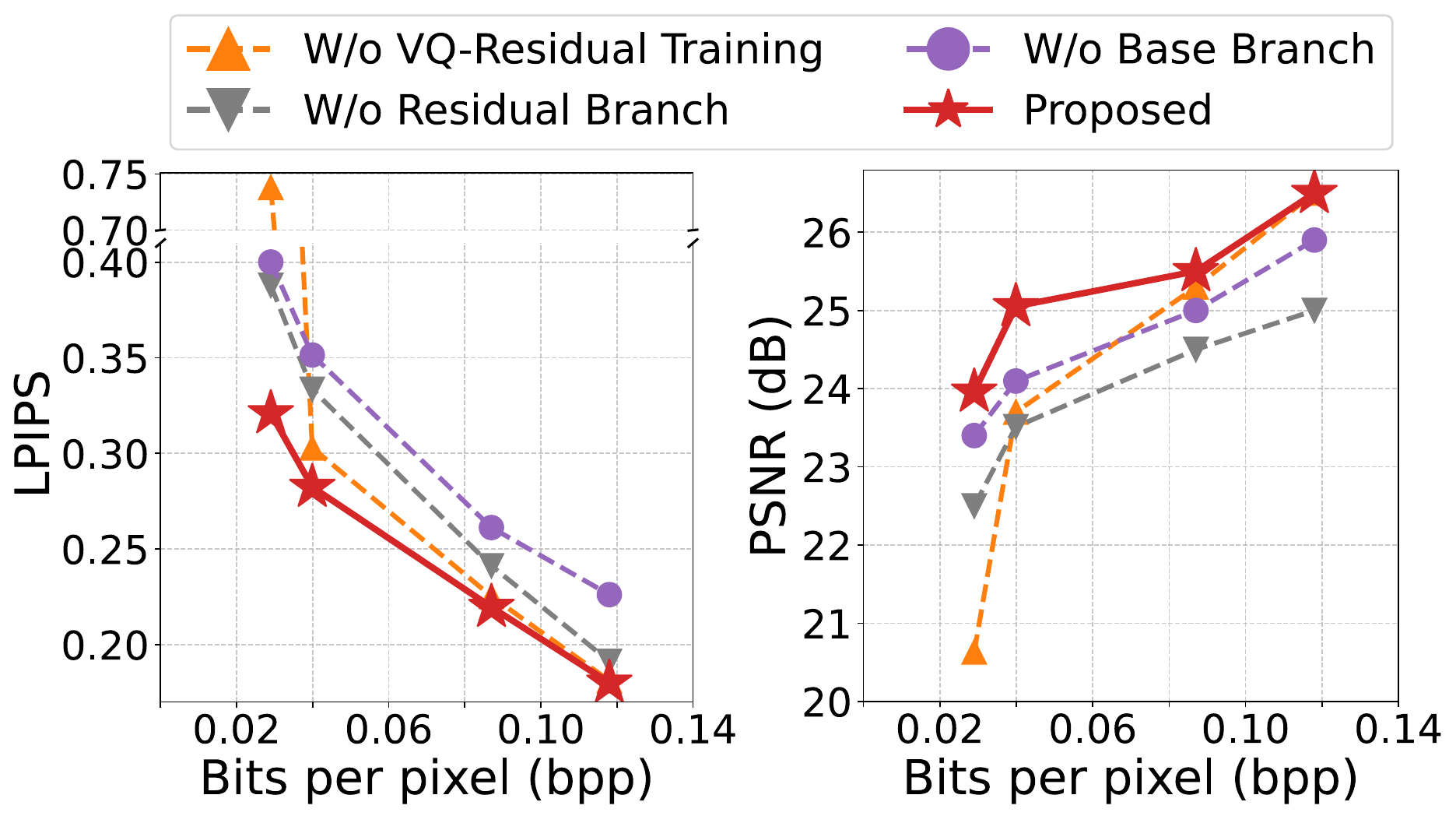}
\end{center}
\vspace{-1.0em}
\caption{Ablation study of Residual Training Branch on Kodak dataset.}
\vspace{-1.0em}
\label{fig:8}

\end{figure}

To validate the effectiveness of each component in our architecture, we conduct an ablation study on the Kodak dataset, as shown in Fig.~\ref{fig:8}. Specifically, we evaluate the impact of the following components. (1) VQ-Residual Training: Disabling this component significantly degrades performance, particularly at lower bitrates, where LPIPS increases sharply. This demonstrates the importance of structurally coherent residual representation for perceptual quality. (2) Residual Branch: Removing the residual branch leads to consistently worse LPIPS scores across all bitrates, confirming that the residual pathway plays a crucial role in reconstructing high-level structures that are often lost during compression. (3) Base Branch: Excluding the base branch also results in degraded performance, especially in mid-to-high bitrate regimes. This shows that perceptual refinement through denoising is essential for enhancing detail and visual fidelity. Our proposed full model achieves the best LPIPS across all bitrates, verifying that the synergistic combination of both residual and base branches, along with VQ-Residual training, is critical for high-quality image reconstruction under extreme compression.

As shown in Table~\ref{tab:step}, the proposed 2-step diffusion training method outperforms both the naive single-step denoising and the costly 15-step approach, achieving superior performance in terms of both perceptual quality and distortion. Notably, while the 2-step model provides moderate improvements in distortion (PSNR, MS-SSIM), it compromises perceptual fidelity, as reflected in increased LPIPS and DISTS. Our method mitigates this trade-off by unifying distortion-aware and perceptual objectives within a single inference step.

\section{Conclusion}


In this work, we proposed a single-step diffusion method for perceptual image compression under ultra-low bitrates.
Our framework combines VQ-Residual training for accurate detail reconstruction with rate-aware noise modulation. Coupled with the inherently low variance of VQ-based bitrates, this ensures both predictable bitrates and high-quality reconstructions  with a fixed single-step process.
Experiments show that our method delivers competitive perceptual fidelity while achieving over 50× faster decoding than prior diffusion-based codecs, highlighting its practicality for real-world, bandwidth-constrained applications.

{
    \small
    \bibliographystyle{ieeenat_fullname}
    \bibliography{egbib}

\begin{thebibliography}{44}
\providecommand{\natexlab}[1]{#1}
\providecommand{\url}[1]{\texttt{#1}}
\expandafter\ifx\csname urlstyle\endcsname\relax
  \providecommand{\doi}[1]{doi: #1}\else
  \providecommand{\doi}{doi: \begingroup \urlstyle{rm}\Url}\fi

\bibitem[Agustsson et~al.(2019)Agustsson, Tschannen, Mentzer, Timofte, and Gool]{agustsson2019generative}
Eirikur Agustsson, Michael Tschannen, Fabian Mentzer, Radu Timofte, and Luc~Van Gool.
\newblock Generative adversarial networks for extreme learned image compression.
\newblock In \emph{Proceedings of the IEEE/CVF international conference on computer vision}, pages 221--231, 2019.

\bibitem[Ball{\'e} et~al.(2017)Ball{\'e}, Laparra, and Simoncelli]{ballelic}
Johannes Ball{\'e}, Valero Laparra, and Eero~P Simoncelli.
\newblock End-to-end optimized image compression.
\newblock In \emph{International Conference on Learning Representations}, 2017.

\bibitem[Ball{\'e} et~al.(2018)Ball{\'e}, Minnen, Singh, Hwang, and Johnston]{hyperprior}
Johannes Ball{\'e}, David Minnen, Saurabh Singh, Sung~Jin Hwang, and Nick Johnston.
\newblock Variational image compression with a scale hyperprior.
\newblock In \emph{International Conference on Learning Representations}, 2018.

\bibitem[Bellard()]{bpg}
Fabrice Bellard.
\newblock {BPG Image Format}.
\newblock \url{https://bellard.org/bpg/}.
\newblock Accessed: 2025-05-15.

\bibitem[Bjontegaard(2001)]{bd-rate}
Gisle Bjontegaard.
\newblock Calculation of average psnr differences between rd-curves.
\newblock \emph{ITU SG16 Doc. VCEG-M33}, 2001.

\bibitem[Bross et~al.(2021)Bross, Wang, Ye, Liu, Chen, Sullivan, and Ohm]{vvc}
Benjamin Bross, Ye-Kui Wang, Yan Ye, Shan Liu, Jianle Chen, Gary~J. Sullivan, and Jens-Rainer Ohm.
\newblock Overview of the versatile video coding (vvc) standard and its applications.
\newblock \emph{IEEE Transactions on Circuits and Systems for Video Technology}, 31\penalty0 (10):\penalty0 3736--3764, 2021.

\bibitem[Careil et~al.(2024)Careil, Muckley, Verbeek, and Lathuili{\`e}re]{perco}
Marlene Careil, Matthew~J. Muckley, Jakob Verbeek, and St{\'e}phane Lathuili{\`e}re.
\newblock Towards image compression with perfect realism at ultra-low bitrates.
\newblock In \emph{The Twelfth International Conference on Learning Representations}, 2024.

\bibitem[Cheng et~al.(2020)Cheng, Sun, Takeuchi, and Katto]{cheng2020}
Zhengxue Cheng, Heming Sun, Masaru Takeuchi, and Jiro Katto.
\newblock Learned image compression with discretized gaussian mixture likelihoods and attention modules.
\newblock In \emph{Proceedings of the IEEE Conference on Computer Vision and Pattern Recognition (CVPR)}, 2020.

\bibitem[Christopoulos et~al.(2000)Christopoulos, Skodras, and Ebrahimi]{jpeg2000}
C. Christopoulos, A. Skodras, and T. Ebrahimi.
\newblock The jpeg2000 still image coding system: an overview.
\newblock \emph{IEEE Transactions on Consumer Electronics}, 46\penalty0 (4):\penalty0 1103--1127, 2000.

\bibitem[Deng et~al.(2009)Deng, Dong, Socher, Li, Li, and Fei-Fei]{imagenet}
Jia Deng, Wei Dong, Richard Socher, Li-Jia Li, Kai Li, and Li Fei-Fei.
\newblock Imagenet: A large-scale hierarchical image database.
\newblock In \emph{2009 IEEE Conference on Computer Vision and Pattern Recognition}, pages 248--255, 2009.

\bibitem[Ding et~al.(2020)Ding, Ma, Wang, and Simoncelli]{dists}
Keyan Ding, Kede Ma, Shiqi Wang, and Eero~P Simoncelli.
\newblock Image quality assessment: Unifying structure and texture similarity.
\newblock \emph{IEEE transactions on pattern analysis and machine intelligence}, 44\penalty0 (5):\penalty0 2567--2581, 2020.

\bibitem[{Eastman Kodak Company}()]{kodak}
{Eastman Kodak Company}.
\newblock {Kodak Lossless True Color Image Suite}.
\newblock \url{http://r0k.us/graphics/kodak/}.

\bibitem[Esser et~al.(2021)Esser, Rombach, and Ommer]{vqgan}
Patrick Esser, Robin Rombach, and Bjorn Ommer.
\newblock Taming transformers for high-resolution image synthesis.
\newblock In \emph{Proceedings of the IEEE/CVF conference on computer vision and pattern recognition}, pages 12873--12883, 2021.

\bibitem[He et~al.(2022)He, Yang, Peng, Ma, Qin, and Wang]{elic}
Dailan He, Ziming Yang, Weikun Peng, Rui Ma, Hongwei Qin, and Yan Wang.
\newblock Elic: Efficient learned image compression with unevenly grouped space-channel contextual adaptive coding.
\newblock In \emph{Proceedings of the IEEE/CVF Conference on Computer Vision and Pattern Recognition}, pages 5718--5727, 2022.

\bibitem[Ho et~al.(2020)Ho, Jain, and Abbeel]{ddpm}
Jonathan Ho, Ajay Jain, and Pieter Abbeel.
\newblock Denoising diffusion probabilistic models.
\newblock \emph{Advances in neural information processing systems}, 33:\penalty0 6840--6851, 2020.

\bibitem[Hoogeboom et~al.(2023)Hoogeboom, Agustsson, Mentzer, Versari, Toderici, and Theis]{hoogeboom}
Emiel Hoogeboom, Eirikur Agustsson, Fabian Mentzer, Luca Versari, George Toderici, and Lucas Theis.
\newblock High-fidelity image compression with score-based generative models.
\newblock \emph{arXiv preprint arXiv:2305.18231}, 2023.

\bibitem[Iwai et~al.(2021)Iwai, Miyazaki, Sugaya, and Omachi]{fidelity}
Shoma Iwai, Tomo Miyazaki, Yoshihiro Sugaya, and Shinichiro Omachi.
\newblock Fidelity-controllable extreme image compression with generative adversarial networks.
\newblock In \emph{2020 25th International Conference on Pattern Recognition (ICPR)}, pages 8235--8242. IEEE, 2021.

\bibitem[Jia et~al.(2024)Jia, Li, Li, Li, and Lu]{glc}
Zhaoyang Jia, Jiahao Li, Bin Li, Houqiang Li, and Yan Lu.
\newblock Generative latent coding for ultra-low bitrate image compression.
\newblock In \emph{Proceedings of the IEEE/CVF Conference on Computer Vision and Pattern Recognition}, pages 26088--26098, 2024.

\bibitem[Karras et~al.(2022)Karras, Aittala, Aila, and Laine]{edm}
Tero Karras, Miika Aittala, Timo Aila, and Samuli Laine.
\newblock Elucidating the design space of diffusion-based generative models.
\newblock \emph{Advances in neural information processing systems}, 35:\penalty0 26565--26577, 2022.

\bibitem[Li et~al.(2025{\natexlab{a}})Li, Zhou, Wei, Ge, and Jiang]{diffeic}
Zhiyuan Li, Yanhui Zhou, Hao Wei, Chenyang Ge, and Jingwen Jiang.
\newblock Toward extreme image compression with latent feature guidance and diffusion prior.
\newblock \emph{IEEE Transactions on Circuits and Systems for Video Technology}, 35\penalty0 (1):\penalty0 888--899, 2025{\natexlab{a}}.

\bibitem[Li et~al.(2025{\natexlab{b}})Li, Zhou, Wei, Ge, and Mian]{rdeic}
Zhiyuan Li, Yanhui Zhou, Hao Wei, Chenyang Ge, and Ajmal Mian.
\newblock Rdeic: Accelerating diffusion-based extreme image compression with relay residual diffusion, 2025{\natexlab{b}}.

\bibitem[Lu et~al.(2022)Lu, Zhou, Bao, Chen, Li, and Zhu]{dpm}
Cheng Lu, Yuhao Zhou, Fan Bao, Jianfei Chen, Chongxuan Li, and Jun Zhu.
\newblock Dpm-solver: A fast ode solver for diffusion probabilistic model sampling in around 10 steps.
\newblock \emph{Advances in Neural Information Processing Systems}, 35:\penalty0 5775--5787, 2022.

\bibitem[Lu et~al.(2025)Lu, Li, Wang, Wang, and Jiang]{hdc}
Lei Lu, Yize Li, Yanzhi Wang, Wei Wang, and Wei Jiang.
\newblock Hdcompression: Hybrid-diffusion image compression for ultra-low bitrates.
\newblock \emph{arXiv preprint arXiv:2502.07160}, 2025.

\bibitem[Mentzer et~al.(2020)Mentzer, Toderici, Tschannen, and Agustsson]{hific}
Fabian Mentzer, George~D Toderici, Michael Tschannen, and Eirikur Agustsson.
\newblock High-fidelity generative image compression.
\newblock \emph{Advances in neural information processing systems}, 33:\penalty0 11913--11924, 2020.

\bibitem[Minnen et~al.(2018)Minnen, Ball{\'e}, and Toderici]{mbt2018}
David Minnen, Johannes Ball{\'e}, and George~D Toderici.
\newblock Joint autoregressive and hierarchical priors for learned image compression.
\newblock \emph{Advances in neural information processing systems}, 31, 2018.

\bibitem[Muckley et~al.(2023{\natexlab{a}})Muckley, El-Nouby, Ullrich, J{\'e}gou, and Verbeek]{improving}
Matthew~J Muckley, Alaaeldin El-Nouby, Karen Ullrich, Herv{\'e} J{\'e}gou, and Jakob Verbeek.
\newblock Improving statistical fidelity for neural image compression with implicit local likelihood models.
\newblock In \emph{International Conference on Machine Learning}, pages 25426--25443. PMLR, 2023{\natexlab{a}}.

\bibitem[Muckley et~al.(2023{\natexlab{b}})Muckley, El-Nouby, Ullrich, J{\'e}gou, and Verbeek]{msillm}
Matthew~J Muckley, Alaaeldin El-Nouby, Karen Ullrich, Herv{\'e} J{\'e}gou, and Jakob Verbeek.
\newblock Improving statistical fidelity for neural image compression with implicit local likelihood models.
\newblock In \emph{International Conference on Machine Learning}, pages 25426--25443. PMLR, 2023{\natexlab{b}}.

\bibitem[Raman et~al.(2020)Raman, Ramesh, Naganoor, Dash, Kumaravelu, and Lee]{compressnet}
Suraj~Kiran Raman, Aditya Ramesh, Vijayakrishna Naganoor, Shubham Dash, Giridharan Kumaravelu, and Honglak Lee.
\newblock Compressnet: Generative compression at extremely low bitrates.
\newblock In \emph{Proceedings of the IEEE/CVF Winter Conference on Applications of Computer Vision}, pages 2325--2333, 2020.

\bibitem[Relic et~al.(2024)Relic, Azevedo, Gross, and Schroers]{disney}
Lucas Relic, Roberto Azevedo, Markus Gross, and Christopher Schroers.
\newblock Lossy image compression with foundation diffusion models.
\newblock In \emph{European Conference on Computer Vision}, pages 303--319. Springer, 2024.

\bibitem[Rombach et~al.(2022)Rombach, Blattmann, Lorenz, Esser, and Ommer]{ldm}
Robin Rombach, Andreas Blattmann, Dominik Lorenz, Patrick Esser, and Bj{\"o}rn Ommer.
\newblock High-resolution image synthesis with latent diffusion models.
\newblock In \emph{Proceedings of the IEEE/CVF conference on computer vision and pattern recognition}, pages 10684--10695, 2022.

\bibitem[Salimans and Ho(2022)]{salimans}
Tim Salimans and Jonathan Ho.
\newblock Progressive distillation for fast sampling of diffusion models.
\newblock \emph{arXiv preprint arXiv:2202.00512}, 2022.

\bibitem[Song et~al.(2020)Song, Meng, and Ermon]{ddim}
Jiaming Song, Chenlin Meng, and Stefano Ermon.
\newblock Denoising diffusion implicit models.
\newblock \emph{arXiv preprint arXiv:2010.02502}, 2020.

\bibitem[Song et~al.(2023)Song, Dhariwal, Chen, and Sutskever]{consistency}
Yang Song, Prafulla Dhariwal, Mark Chen, and Ilya Sutskever.
\newblock Consistency models.
\newblock 2023.

\bibitem[Sullivan et~al.(2012)Sullivan, Ohm, Han, and Wiegand]{hevc}
Gary~J. Sullivan, Jens-Rainer Ohm, Woo-Jin Han, and Thomas Wiegand.
\newblock Overview of the high efficiency video coding (hevc) standard.
\newblock \emph{IEEE Transactions on Circuits and Systems for Video Technology}, 22\penalty0 (12):\penalty0 1649--1668, 2012.

\bibitem[Theis et~al.(2022)Theis, Salimans, Hoffman, and Mentzer]{diffc}
Lucas Theis, Tim Salimans, Matthew~D Hoffman, and Fabian Mentzer.
\newblock Lossy compression with gaussian diffusion.
\newblock \emph{arXiv preprint arXiv:2206.08889}, 2022.

\bibitem[Toderici et~al.(2020)Toderici, Theis, Johnston, Agustsson, Mentzer, Ballé, Shi, and Timofte]{clic2020}
George Toderici, Lucas Theis, Nick Johnston, Eirikur Agustsson, Fabian Mentzer, Johannes Ballé, Wenzhe Shi, and Radu Timofte.
\newblock {CLIC 2020: Challenge on Learned Image Compression}.
\newblock \url{http://www.compression.cc}, 2020.

\bibitem[Van Den~Oord et~al.(2017)Van Den~Oord, Vinyals, et~al.]{vqvae}
Aaron Van Den~Oord, Oriol Vinyals, et~al.
\newblock Neural discrete representation learning.
\newblock \emph{Advances in neural information processing systems}, 30, 2017.

\bibitem[Wallace(1992)]{jpeg}
G.K. Wallace.
\newblock The jpeg still picture compression standard.
\newblock \emph{IEEE Transactions on Consumer Electronics}, 38\penalty0 (1):\penalty0 xviii--xxxiv, 1992.

\bibitem[Wang et~al.(2003)Wang, Simoncelli, and Bovik]{ms-ssim}
Zhou Wang, Eero~P Simoncelli, and Alan~C Bovik.
\newblock Multiscale structural similarity for image quality assessment.
\newblock In \emph{The Thrity-Seventh Asilomar Conference on Signals, Systems \& Computers, 2003}, pages 1398--1402. Ieee, 2003.

\bibitem[Xia et~al.(2025)Xia, Zhou, Wang, An, Wang, Wang, and Chen]{diffpc}
Yichong Xia, Yimin Zhou, Jinpeng Wang, Baoyi An, Haoqian Wang, Yaowei Wang, and Bin Chen.
\newblock Diff{PC}: Diffusion-based high perceptual fidelity image compression with semantic refinement.
\newblock In \emph{The Thirteenth International Conference on Learning Representations}, 2025.

\bibitem[Yang and Mandt(2023)]{cdc}
Ruihan Yang and Stephan Mandt.
\newblock Lossy image compression with conditional diffusion models.
\newblock \emph{Advances in Neural Information Processing Systems}, 36:\penalty0 64971--64995, 2023.

\bibitem[Yin et~al.(2024)Yin, Gharbi, Zhang, Shechtman, Durand, Freeman, and Park]{dmd}
Tianwei Yin, Micha{\"e}l Gharbi, Richard Zhang, Eli Shechtman, Fredo Durand, William~T Freeman, and Taesung Park.
\newblock One-step diffusion with distribution matching distillation.
\newblock In \emph{Proceedings of the IEEE/CVF conference on computer vision and pattern recognition}, pages 6613--6623, 2024.

\bibitem[Yue et~al.(2023)Yue, Wang, and Loy]{resshift}
Zongsheng Yue, Jianyi Wang, and Chen~Change Loy.
\newblock Resshift: Efficient diffusion model for image super-resolution by residual shifting.
\newblock \emph{Advances in Neural Information Processing Systems}, 36:\penalty0 13294--13307, 2023.

\bibitem[Zhang et~al.(2018)Zhang, Isola, Efros, Shechtman, and Wang]{lpips}
Richard Zhang, Phillip Isola, Alexei~A Efros, Eli Shechtman, and Oliver Wang.
\newblock The unreasonable effectiveness of deep features as a perceptual metric.
\newblock In \emph{Proceedings of the IEEE conference on computer vision and pattern recognition}, pages 586--595, 2018.

\end{thebibliography}
}

\end{document}